\documentclass{article}

\usepackage{microtype}
\usepackage{graphicx}
\usepackage{subfigure}
\usepackage{booktabs} % for professional tables

\usepackage{hyperref}

\usepackage[accepted]{icml2025}

\usepackage{amsmath}
\usepackage{amssymb}
\usepackage{mathtools}
\usepackage{amsthm}

\usepackage[capitalize,noabbrev]{cleveref}

\theoremstyle{plain}

\theoremstyle{definition}

\theoremstyle{remark}

\usepackage[textsize=tiny]{todonotes}

\usepackage[acronym]{glossaries}
\usepackage{caption} 
\usepackage[T1]{fontenc}
\usepackage[shortlabels]{enumitem}
\setlist[enumerate]{nosep}

\makeglossaries

\newacronym{MAS}{MAS}{Multi-Agent Systems}
\newacronym{LLM}{LLM}{Large Language Models}
\newacronym{AI}{AI}{Artificial Intelligence}
\newacronym{RAG}{RAG}{Retrieval-Augmented Generation}

\icmltitlerunning{Position: Towards a Responsible LLM-empowered Multi-Agent Systems}

\begin{document}

\twocolumn[
\icmltitle{Position: Towards a Responsible LLM-empowered Multi-Agent Systems}



\icmlsetsymbol{equal}{*}

\begin{icmlauthorlist}
% \icmlauthor{Anonymous Authors}{yyy}
% \icmlauthor{Firstname2 Lastname2}{equal,yyy,comp}
% 
\icmlauthor{Jinwei Hu}{uol}
\icmlauthor{Yi Dong}{uol}
\icmlauthor{Shuang Ao}{uos}
\icmlauthor{Zhuoyun Li}{uol}
\icmlauthor{Boxuan Wang}{uol}
\icmlauthor{Lokesh Singh}{uos}
\icmlauthor{Guangliang Cheng}{uol}
\icmlauthor{Sarvapali D. Ramchurn}{uos}
\icmlauthor{Xiaowei Huang}{uol}
\end{icmlauthorlist}

\icmlaffiliation{uol}{Department of Computer Science, University of Liverpool, Liverpool, UK}
\icmlaffiliation{uos}{Department of Electronics and Computer Science, University of Southampton, Southampton, UK}
% \icmlaffiliation{yyy}{Company Name, Location, Country}

\icmlcorrespondingauthor{Yi Dong}{yi.dong@liverpool.ac.uk}

\icmlkeywords{Machine Learning, ICML}

\vskip 0.3in
]



\printAffiliationsAndNotice{}  % leave blank if no need to mention equal contribution

\begin{abstract}
% Multi-Agent Systems (MAS) involving autonomous agents from software to humans increasingly leverage Large Language Models (LLMs) to enhance decision-making capabilities. Tools like LangChain and Retrieval Augmented Generation (RAG) have revolutionized LLM functionalities, enabling deeper integration with MAS by providing access to extensive knowledge bases and sophisticated reasoning abilities. 
% The rise of Agent AI and Large Language Models (LLM) empowered Multi-Agent Systems (MAS) has underscored the need for responsible and dependable system operation. 
% Tools like LangChain and Retrieval Augmented Generation have expanded LLM capabilities, enabling deeper integration into MAS through extensive knowledge bases and advanced reasoning abilities. 
% This integration, however, introduces challenges such as the unpredictability of LLM agents and the accumulation of uncertainties, potentially compromising system stability. 

The rise of Agent AI and Large Language Model-powered Multi-Agent Systems (LLM-MAS) has underscored the need for responsible and dependable system operation. Tools like LangChain and Retrieval-Augmented Generation have expanded LLM capabilities, enabling deeper integration into MAS through enhanced knowledge retrieval and reasoning. However, these advancements introduce critical challenges: LLM agents exhibit inherent unpredictability, and uncertainties in their outputs can compound across interactions, threatening system stability. To address these risks, \textbf{a human-centered design approach with active dynamic moderation is essential}. Such an approach enhances traditional passive oversight by facilitating coherent inter-agent communication and effective system governance, allowing MAS to achieve desired outcomes more efficiently.

\end{abstract}

\section{Introduction}

\gls{MAS} represent a critical area of research in decision-making, where multiple autonomous agents\footnote{$Agent$ could be software agent, robotics agent, embodied agent or human agent.} interact within a defined environment to achieve individual or collective goals. In the rapidly evolving landscape of \gls{LLM}
, tools like LangChain \cite{topsakal2023creating} have begun to revolutionize the way we interact with \gls{LLM}, enabling a programming-like interface for sculpting application-specific interactions. Furthermore, technologies such as \gls{RAG} \cite{lewis2020retrieval} enhance \gls{LLM} capabilities by allowing them to access external databases and even other tools, and therefore broadening their operational horizon. 
The integration of \gls{LLM} into \gls{MAS} has further extended the decision-making capabilities, providing a huge knowledge base and advanced reasoning abilities that significantly enhance efficiency beyond what is achievable by human efforts alone.
However, this integration introduces new challenges that are absent in traditional MAS setups.

A core challenge in LLM-MAS intrinsically is achieving enhanced mutual understanding among agents. Unlike traditional MAS with predefined protocols ensuring \textit{deterministic behaviours}, LLM-based agents, trained on diverse datasets, exhibit emergent and \textit{unpredictable behaviours}. This unpredictability create a need for quantifiable mechanisms, such as trust metrics, to facilitate and verify effective agreement among agents. Without such mechanisms, agents may struggle to interpret or align with one another's actions.

Beyond the internal challenges of agent interaction, LLM-MAS face external challenges related to uncertainty propagation. As these systems grow in complexity, the inherent uncertainties of individual LLM agents can accumulate and cascade through the network \cite{10.5555/3692070.3692731}, potentially compromising system correctness and stability. This challenge becomes particularly salient when considering the lifecycle of LLM-MAS, where uncertainties must be quantified and managed at both individual agent-level and the system level.

To address these challenges while harnessing the powerful knowledge representation and reasoning capabilities of LLM, A human-centered design approach is essential. This approach incorporates active dynamic moderation as a core component of LLM-MAS, moving beyond traditional passive oversight.
The moderator plays a critical role in system governance, engaging in collaborative decision-making, providing high-level perspectives to LLM agents, implementing real-time intervention protocols, and steering the system toward desired outcomes.

In this paper, we posit that:

\begin{enumerate}
    \item \textbf{Agents must 
    %be able to 
    "understand" one another}, necessitating quantifiable metrics with probabilistic guarantees to assess inter-agent \textit{agreement}\footnote{Different from \textit{alignment} that focuses on individual agent's conformity to external objectives (generally ethical value, human intentions, or specific requirements), \textbf{agreement} emphasizes both system-level \textit{behavioural coherence} and inter-agent \textit{mutual understanding} (e.g., coordinated outputs, decisions, strategies, and unified semantic interpretations across agents).}  under uncertainty.%necessitating the creation of mechanisms to quantify such %agreement via guaranteed uncertainty metric, 
    %for instance, e.g. trust scores.
    % \item There is a necessity to manage uncertainty for LLM-MAS.
    \item \textbf{Robust mechanisms for uncertainty quantification and management are essential}, operating at both the agent and system levels to ensure control throughout the lifecycle.
    % mitigate the negative aspects introduced by LLMs, such as their inherent biases and the potential for unpredictable behaviours.
    \item \textbf{A human-centered system-level moderator is needed} to oversee, participate in, and guide the MAS, seamlessly integrating human oversight with automated processes.
\end{enumerate}

The goal of this paper is to review the state-of-the-art vulnerabilities and challenges in existing LLM-MAS (Section 2). Then, current solutions for the internal (Agreement, Section 3) and external (Uncertainty, Section 4) challenges of responsible LLM-MAS are discussed. Finally, Section 5 explores potential 
%solutions
research directions to address these challenges and achieve responsible LLM-MAS. %  detailed internal challenges propose a potential framework for responsible LLM-MAS (Section 4) present technical challenges in building an LLM-MAS, and discuss several issues regarding the systematic design of an LLM-MAS (Section 5).

\section{Challenges in Existing LLM-MAS} 
In this section, we first conduct a comprehensive examination of the intrinsic challenges and systemic vulnerabilities in LLM-MAS, followed by our perspectives and potential solutions to address these issues.
%e.g., knowledge drift, misinformation propagation, conflicting agreements among agents, inherent behaviors like hallucination, collusion, and emerging security threats. We also highlight the challenges in evaluating system-level uncertainties within LLM-MAS, 
% underscoring the necessity for designing a dedicated mechanism for, responsible LLM-MAS.
\subsection{Knowledge Drift \& Misinformation Propagation} \label{Knowledge Drift and Misinformation Propagation}
Unlike traditional MAS with explicitly programmed goals, LLM-MAS faces unique challenges such as ``knowledge drift" and ``misinformed perspective propagation", stemming from the inherent variability and probabilistic nature in natural language processing \cite{fastowski2024understanding,xu-etal-2024-earth,wang2024rethinking}. These challenges are particularly pronounced in collaborative reasoning 
%and debate 
tasks, where phenomena like the conformity effect and authoritative bias lead agents to align with wrong consensus or defer to perceived authority, amplifying reasoning errors and distorting knowledge bases—even some agents initially hold correct viewpoints \cite{zhang2024exploring}. For instance, in multi-agent debates, an agent with a partially flawed understanding may generate persuasive yet erroneous rationales, potentially impacting others and collectively diverting the reasoning path from accurate solutions \cite{breum2024persuasive}.

Additionally, LLM agents exhibit a tendency for ``cognitive bias expansion," wherein, unlike humans who compress and filter information, they amplify and propagate errors, further exacerbating knowledge drift and collective reasoning inaccuracies \cite{liu2024exploring}. Existing approaches, such as prompt engineering \cite{fernando2024promptbreeder}, the use of LLM agents as judge to arbitrate and refine reasoning \cite{zheng2023judging,chan2024chateval}, and ``human-in-the-loop" intervention \cite{triem2024tipping}, attempt to address these issues. However, prompt engineering often lacks scalability and struggles with context-specific biases, while human intervention is labour-intensive and impractical for large-scale systems. Moreover, judge agents, being LLM-based themselves, are susceptible to similar biases and can unintentionally reinforce reasoning errors, leaving knowledge drift a persistent challenge \cite{wang2024rethinking}. In contrast, methods integrating uncertainty have shown improved performance; however, their reliance on open-source LLMs, sensitivity to decision-making strategies, and lack of theoretical assurances limit their applicability to proprietary models and complex multi-agent real-world scenarios \cite{yoffe2024debunc,yang2024confidence,DBLP:journals/corr/abs-2407-11282}. These limitations underscore the need for a paradigm shift in LLM-empowered MAS design, demanding a framework that \textit{leverages quantifiable uncertainty to mitigate knowledge drift and misinformation propagation while providing robust theoretical guarantees for the whole system.}
 
\textbf{Our Perspective:} Addressing aforementioned issues in LLM-MAS requires a transition from current heuristic solutions to principled system architectures with \textit{provable guarantees}, particularly to ensure reliable knowledge agreement \cite{bensalem2023indeed}. Different from existing approaches based on heuristic mechanisms, we advocate a probabilistic-centric system architecture that fundamentally integrates uncertainty quantification and propagation mechanisms into its core operational principles to ensure consistent knowledge alignment across whole agent network instead of focusing on individual agents. 
% This architecture should be underpinned by both \textit{statistical bounds} (similar to probably approximately correct (PAC) learning guarantees) \cite{kearns1994introduction,brand2023parameterized,mocanu2023knowledge} and \textit{deterministic bounds} \cite{huang2017safety,kumar2023certifying} that certify the robustness of aligned collective knowledge under various operating conditions. 
Specifically, we propose that future LLM-MAS should: (1) implement rigorous probabilistic frameworks for quantifying and propagating uncertainty in inter-agent communications to maintain agreement consistency, (2) establish formal verification mechanisms that provide certified bounds (either \textit{statistical} or \textit{deterministic bounds}) on the probabilities of knowledge corruption and drift \cite{zhang2024fusion}, and (3) develop scalable certification procedures with automated assurance cases for efficient agreement verification \cite{wang2023computer}. For instance,  conformal prediction-style guarantees have been used to ensure collective decisions align with a specified confidence level while quantifying individual agent uncertainties \cite{wang2024probabilistically,vishwakarma2024improving}. 
% This principled design should facilitate mathematical verification of system-wide alignment properties, offering theoretical guarantees for preserving knowledge integrity and consistency, even as systems scale to larger agent populations and increasingly complex interaction patterns. By advancing this architectural framework, we transition from empirical, post-hoc adjustments to verifiable assurances of reliable knowledge alignment in LLM-empowered multi-agent systems.

\subsection{Conflicting Agreement}
Conflicts in LLM-MAS normally arise from objective misalignment and knowledge asymmetry \cite{phelps2023models}. At the objective level, conflicts stem from differing task criteria or requirement interpretations. For example, in collaborative task planning, agents may adopt competing interpretations of the same high-level goal (typically performance vs. safety), resulting in divergent execution strategies, particularly in scenarios requiring complex trade-offs \cite{tessier2005conflicting}. Knowledge-based conflicts emerge from different reasoning paths and knowledge sources, where agents may construct different mental models or reach contradictory conclusions despite identical initial information \cite{wang2024astute}. This is evident in RAG-enhanced systems where variations in chain-of-thought reasoning and retrieved knowledge lead to inconsistent understanding across temporal and domain-specific contexts \cite{DBLP:journals/corr/abs-2407-07791}. The probabilistic nature of LLMs, coupled with inherent semantic ambiguities in natural language, amplifies the effect of knowledge misalignment. For instance, in an autonomous driving scenario, when one agent issues an alert such as ``slow down due to road conditions," different agents might interpret this message differently, leading to varying implementations of the slowdown \cite{yang2024llmdrive}. While LLMs as agents offer significant advantages
%over traditional approaches
, how do we address the unique conflicts they introduce, posing a new dilemma? That is, \textit{we must determine whether integrating LLMs into MAS can prevent conflicts from inherent knowledge ambiguities in LLM and produce outcomes aligned with our expectations.}

\textbf{Our Perspective:} Current approaches rely mainly on ad-hoc solutions \cite{bhatia2020preference,liu2024autonomous,din2024ontology}, which lack robust mechanisms to quantify and validate uncertainties in decision-making within LLM-MAS, potentially masking conflicts when agents operate with imperfectly alignment levels, easy to allow over-confident yet unreliable decisions \cite{rodriguez2023good}.
% : preference learning for objective conflicts \cite{bhatia2020preference}, consensus mechanisms for knowledge reconciliation \cite{chen2023reconcile,liu2024autonomous}, and prompt engineering for semantic ambiguity \cite{sikha2023mastering,din2024ontology}. However, these methods lack systematic guarantees - Preference learning struggles with dynamic trade-offs and is primarily effective in single agent, consensus mechanisms is sensitive to errors and attacks like hallucination can exist in their shared memory, and prompt engineering has low scalability. Crucially, these approaches lack robust mechanisms to quantify and validate uncertainties in decision-making within LLM-MAS, potentially masking conflicts when agents operate with imperfectly alignment levels, easy to allow over-confident yet unreliable decisions \cite{rodriguez2023good}.
In contrast, we advocate for a principled, theory-driven framework that extends the classical Belief-Desire-Intention (BDI) architecture with guaranteed hierarchical mechanisms for conflict resolution \cite{fischer1995pragmatic}. 
Specifically, the belief layer uses formal verification to standardize 
% ensure consistent 
interpretation of 
% general 
ambiguous
instructions. 
% across agents, standardizing how LLM agents understand ambiguous instructions like ``move cautiously" to construct the initial belief. 
The knowledge layer, extending desire, utilizes probabilistic belief updating (e.g. Conformal Bayesian Inference \cite{fong2021conformal}) 
% where agents attach uncertainty quantification to their beliefs and update them through probabilistic estimation like Conformal Bayesian Inference \cite{fong2021conformal}, enabling principled 
to weight conflicting information based on source reliability and contextual relevance.
% —critical in scenarios like multi-robot navigation where agents must reconcile conflicting environmental observations. 
The objective layer as intention layer, leverages uncertainty-aware multi-criteria decision theory 
% with uncertainty-aware utility functions, 
to explicitly modelling objective priorities and constraints 
for adaptive trade-offs 
% management 
in complex 
% resource allocation 
scenarios. 
% Modelling uncertainty explicitly at each BDI-aligned layer through probabilistic reasoning frameworks, therefore, enables agents to make more reliable decisions while maintaining calibrated confidence levels \cite{ichida2024bdi}. 
This hierarchical design can be augmented by causal reasoning frameworks for preemptive conflict identification \cite{zeng2022multi}. 
\textit{We view conflicts not as anomalies to be eliminated, but as inherent system features requiring dedicated management mechanisms with theoretical foundations.}

\subsection{Inherent Behaviours \& Potential Threats}
\subsubsection{Hallucination}
Hallucination, defined as the generation of fluent yet factually incorrect information, poses more severe systemic risks in multi-agent settings \cite{ji2023survey}. The inherent uncertainty in LLM outputs, driven by their tendency toward overconfident responses, is especially problematic in multi-agent coordination \cite{huang2023survey}. In such scenarios, hallucinated information from one agent can be treated as valid input by others, creating a propagation cycle as mentioned in section \ref{Knowledge Drift and Misinformation Propagation} where false content is not only transmitted but also reinforced through subsequent agent interactions. This vulnerability becomes especially concerning when adversaries can exploit it for persuasive manipulation or collusive behaviours, \textit{transforming an individual agent's uncertainty into a system-wide vulnerability.}

\subsubsection{Collusion}
% The intricate interactions within LLM-MAS present a significant risk of collusion, posing substantial challenges to ethics, fairness, privacy and transparency 
%LLM-MAS face significant 
Collusion is another potential risk, arising both from inter-agent communication and emergent behaviour within individual agents' internal mechanisms \cite{huang2024survey}. For instance, research has demonstrated that LLM agents in Cournot competition can engage in implicit collusion, such as covert market division without explicit coordination, thereby evading detection \cite{wu2024shall,lin2024strategic}.
% Such behaviors can spontaneously occur  
% allowing them to 
Furthermore, 
% agents may exploit implicit 
semantic cues or steganographic techniques further support collusive behaviours, \textit{making them hard to identify and easily exploitable by adversaries} \cite{motwani2024secret}.
% to align decisions surreptitiously, making the identification of collusion even more challenging \cite{motwani2024secret}. This inherent collusive potential can be easily exploited by adversaries as an attack method, enabling them to manipulate the system to cause harm or achieve malicious objectives \cite{ugur2021manipulator}. 
%The opaqueness of 
LLM's opaqueness further exacerbates the issue, as their outputs are often contextually plausible, effectively obscuring the underlying collusive dynamics.
\subsubsection{Data Poisoning \& Jailbreaking Attack}
% Data poisoning and jailbreaking attacks introduce vulnerabilities through communication, contaminated knowledge retrieval, and manipulated context windows in LLM-MAS \cite{das2024security}. 
% Unlike conventional MAS where poisoning primarily targets the training process, LLM-empowered systems face multi-faceted attack vectors through communication, contaminated knowledge retrieval, and manipulated context windows \cite{das2024security}. 
% While jailbreaking traditionally targets individual LLMs to bypass their safety constraints, these compromised outputs in MAS can propagate across the whole agent network \cite{liu2024jailjudge,peng2024jailbreaking}. The natural language communication between agents expands the attack surface, allowing adversaries to exploit LLMs' context sensitivity through subtle linguistic manipulations and safety-bypassing prompts, while RAG mechanisms introduce additional vulnerabilities through poisoned external knowledge bases \cite{chen2024agentpoison}. 
Data poisoning and jailbreaking attacks introduce significant vulnerabilities in LLM-MAS by exploiting communication channels, contaminated knowledge retrieval, and manipulated context windows \cite{das2024security}. Unlike conventional MAS, where poisoning typically targets the training phase, LLM-MAS faces expanded attack vectors due to its reliance on dynamic interactions and external knowledge \cite{das2024security}. For instance, RAG introduces additional risks as it may unguardedly allow poisoned external knowledge bases to infiltrate the originally intact system \cite{chen2024agentpoison}. Furthermore, natural language communication between agents further amplifies the attack surface, allowing adversaries to exploit LLMs' context sensitivity through subtle linguistic manipulations and safety-bypassing prompts. Jailbreaking, normally aimed at bypassing safety constraints in individual LLMs, becomes more dangerous in LLM-MAS \cite{liu2024jailjudge, peng2024jailbreaking}. The property of misinformation propagation leads to both poisoned and jailbroken information being enhanced through collaborative reasoning, creating cascading security breaches across the system. These adversarial settings \textit{highlight the necessity for 
%real-time monitoring 
utilizing a dedicated run-time mechanisms that can continuously detect and filter potentially compromised data throughout the system's operation}, ensuring information consistency and agreement information across agents during task execution.

\subsubsection{Cyber Threats}
Cyber threats %pose 
also become a significant challenge to LLM-MAS due to their distributed architecture and complex interaction patterns \cite{zeeshan2025large}. Network-level attacks, such as wormhole \cite{ren2024hwmp} and denial-of-service \cite{wen2023secure}, can disrupt temporal consistency and degrade operational performance. The frequent API interactions required for LLM services and inter-agent communication not only expose vulnerabilities in network protocols and authentication mechanisms, but also create performance bottlenecks \cite{wang2024large}. Furthermore, the integration of external knowledge sources introduces more attack targets \cite{gummadi2024enhancing}, highlighting the need for robust cybersecurity measures that \textit{balance protection with system responsiveness, while quantifying the timeliness and completeness of information exchange.}

\textbf{Our perspective:} Current mitigation strategies for these risks, while proven effective for individual LLMs, face limitations when extended to LLM-MAS. Traditional hallucination mitigation techniques like retrieval augmentation \cite{shuster-etal-2021-retrieval-augmentation} and static guardrail \cite{dong2024position} is insufficient when hallucinated content can be reinforced and propagated through inter-agent interactions, as false information can gain credibility through repeated validation \cite{xu-etal-2024-earth}. For collusive behaviours, existing detection mechanisms rely heavily on post-hoc analysis of interaction logs, which fails to meet the real-time intervention requirements of dynamic LLM-MAS applications \cite{bonjour2022information,motwani2024secret}. Similarly, data poisoning and jailbreaking defences primarily focus on robust training and input sanitization at model initialization, becoming inadequate in multi-agent scenarios where compromised information can be injected and propagate through various interaction channels during runtime \cite{wang2022threats}. Traditional cybersecurity measures, such as rule-based firewalls, struggle to address both the uncertainties from dynamic reasoning and the increased communication channels in LLM-MAS \cite{applebaum2016firewall}. Moreover, network-level detection mechanisms have proven less effective against LLM-generated misinformation, as these contents often have more deceptive impact despite semantic equivalence to human-designed attacks \cite{chen2024can}. These approaches, originally designed for \textit{static protection}, cannot effectively handle the \textit{dynamic protection} of knowledge exchange and accumulation in interactive MAS.

We suggest 
% a reconceptualization focused on 
a runtime monitoring and AI provenance framework, enhanced by uncertainty-based governance rules \cite{souza2022workflow,werder2022establishing,xu2022dependency}. This approach emphasizes continuous surveillance of 
% internal representations and 
system behaviours, 
% alongside explicit 
tracking information flow and decision origins. It should integrate provenance chains and uncertainty quantification, then the system can trace and validate information propagation with probabilistic guarantees \cite{shorinwa2024survey}. 
% Achieving this requires theoretical advancements in real-time monitoring, comprehensive provenance tracking, and uncertainty-aware verification, while ensuring efficiency. 
Besides, the framework should enable adaptive monitoring that dynamically adjusts scrutiny based on risk, trust, and reputation, maintaining reliable records of information and decisions \cite{hu2024adaptive}. Also, runtime machine unlearning can 
% be integrated as potential 
remediate 
% measures, allowing LLM agents to selectively forget 
contaminated representations 
% once detected 
\cite{pawelczyk2024incontext}, while 
% To enhance the safety and verifiability of these mechanisms, 
neural-symbolic methods 
% could provide a potential module that 
combine explicit symbolic reasoning (e.g. abductive inference) 
% for explicit monitoring rules (e.g., abductive inference for input-output verification) 
with neural flexibility for safety enhancement \cite{tsamoura2021neural}. By embedding these capabilities within the core architecture, such LLM-MAS should achieve both security and transparency in their operations, providing evidence of system behaviours and their origins during runtime while ensuring robust operation under uncertainty.

% \subsection{Defence} Physical-level, soft level

% \textbf{Our Position:}
\subsection{Evaluation in LLM-MAS}
Evaluating agreement in LLM-MAS shows more difficulties in comparison to a single LLM assessment. The temporal dynamics of LLM agent interactions introduce fundamental evaluation complexities. Capturing the temporal evolution of multi-dimension agreement states, especially under feedback loops and historical dependencies that drive cumulative effects for continuous agreement, remains an open challenge in agent collaboration networks \cite{shen2023large}. For instance, an LLM agent's learning from past interactions may asymmetrically alter its belief alignment and become apparent over extended operational periods \cite{schubert2024context}. Additionally, the probabilistic nature of LLM reasoning means that different sequences of agent interactions can lead to divergent outcomes - for example, in a collaborative planning scenario, having Agent A propose a solution before Agent B might result in a different final strategy compared to when B initiates the planning process, even with identical initial conditions and objectives \cite{yoffe2024debunc}.

Moreover, system-level quantification of agreement faces challenges mostly due to the lack of unified frameworks for aggregating individual agent metrics \cite{guo2024large}. While individual agents might achieve high scores in standard trustworthy dimensions such as toxicity filtering and bias detection, these metrics become insufficient in multi-agent scenarios where agents can reinforce biases through their interactions. Even performance metrics like response efficiency and task completion rates fail to reflect emergent behaviours in collaborative scenarios, where individually optimal responses might lead to collectively suboptimal outcomes, particularly when LLMs inherently have selfish strategies such as maintaining conversational dominance \cite{tennant2024moral}. Notably, \cite{wang2024rethinking} demonstrate that interaction dynamics can lead to worse performance compared to single-agent's solutions, \textit{indicating that the participation of more individually well-performing agents does not necessarily lead to better outcomes.}

\textbf{Our Perspective:} Current approaches to evaluating agreement in LLM-MAS primarily focus on \textit{static measurement and metric extension from single agent to multiagent, overlooking the dynamic evolution of multi-agent agreement during task execution}. Moreover, recent attempts directly use LLMs as dynamic evaluators, but these evaluations still \textit{lack theoretical guarantees} and can be highly \textit{sensitive to subjective factors} like prompt template design \cite{wei2024systematic}. We advocate a learning-based method that can dynamically adapts to the evolving characteristics of agent interactions. 
% Specifically, we propose employing learning 
Using techniques like metric learning \cite{huisman2021survey} or submodular optimization \cite{chen2024less}, it synthesizes global and local evaluation functions, optimizing multi-dimensional agreement metrics based on observed agent behaviours and interaction patterns. 
% Rather than relying on predefined static metrics, t
This approach is able to learn context-aware subspace projections,
% that capture meaningful alignment characteristics across different interaction contexts. Through such multi-criteria learning strategies, the LLM-MAS should be able to jointly optimizes multiple evaluation functions while producing 
enabling probabilistic interpretability of system performance 
% across different criteria 
\cite{liao2023reimagining}, and 
% . For instance, measuring relative metric contributions or establishing confidence intervals for alignment assessments, thus 
providing transparent insight into both overall system agreement and 
% specific aspects of 
trustworthiness.

\section{Agreement in LLM-MAS}
As LLMs become increasingly embedded in agents, LLM-MAS has demonstrated unprecedented capabilities in complex task solving \cite{bubeck2023sparksartificialgeneralintelligence}. This integration necessitates a reconceptualization of system-wide safety and efficiency beyond traditional protocol-based approaches. From a internal perspective of LLM-MAS, the primary objective is to achieve \textit{global agreement}
% \footnote{Different from \textit{alignment} that focuses on individual agent's conformity to external objectives (generally ethical value, human intentions, or specific requirements), \textbf{agreement} emphasizes both system-level \textit{behavioural coherence} and inter-agent \textit{mutual understanding} (e.g., coordinated outputs, decisions, strategies, and unified semantic interpretations across agents).} 
\cite{xu2023reasoninglargelanguagemodels, zhao-etal-2024-electoral} across heterogeneous agents, ensuring both ethical and operational consistency through mutual understanding among all components. Recent advances have reviewed some methods in establishing agreement between agents and human intentions, as well as inter-agent coordination. However, existing studies \cite{kirchner2022researchingalignmentresearchunsupervised,shen2023largelanguagemodelalignment, cao2024scalableautomatedalignmentllms, pan2023automaticallycorrectinglargelanguage, fernandes-etal-2023-bridging} mainly focus on the local agreement for single-agent rather than facilitating global agreement for LLM-MAS.

\subsection{Agent to Human Agreement}
% Social and ethical risks are prior to humans and society, requiring LLMs to maintain: helpful, honest, and harmless (HHH)\cite{askell2021generallanguageassistantlaboratory}. Therefore, achieving human expectations and values is an essential cornerstone for agents equipped with LLM\cite{fu2023improvinglanguagemodelnegotiation}. 
% To reach an agreement with humans, agents need to correctly parse the meaning of natural language, know the tasks or goals set by humans, and understand context and constraints. 
% % Here are research methods corresponding to alignment between human and LLMs-based agents.
% Recent advancements within the field have revealed that methodologies for enhancing system performance can primarily be classified into three categories: reinforcement learning, supervised fine-tuning, and self-improvement. 
For establishing agreement with humans, agents must accurately interpret natural language, grasp assigned tasks or goals, and understand societal constraints. Recent advancements broadly classify these agreement-building methods into three categories: reinforcement learning, supervised fine-tuning, and self-improvement.
\begin{figure}[htbp]
\includegraphics[width=0.46\textwidth]{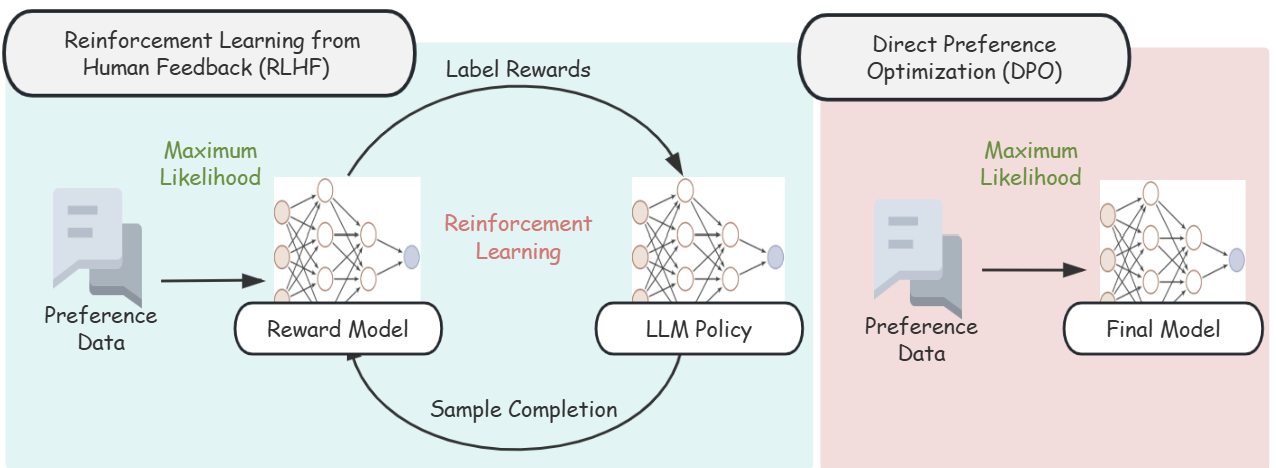} 
    \caption{Framework of Reinforcement Learning} 
    \label{RL} 
\end{figure}

\textbf{Reinforcement Learning}
A most commonly used method to achieve human value agreement is reinforcement learning from human feedback (RLHF) \cite{ouyang2022training, stiennon2020learning, ziegler2019fine}, which is shown in Figure \ref{RL} and includes two steps: train reward models according to collected human feedback data, finetune language models through reinforcement learning (such as a prevalent method Proximal Policy Optimisation (PPO) using policy update \cite{schulman2017proximalpolicyoptimizationalgorithms} ) to achieve agreement. 
Therefore, in \cite{bai2022constitutionalaiharmlessnessai, lee2024rlaifvsrlhfscaling}, human feedback is replaced and compared by off-the-shelf LLMs to save human work on high-quality preference labels. Then RLHF are further enhanced in \cite{glaese2022improvingalignmentdialogueagents, bai2022traininghelpfulharmlessassistant, tan-etal-2023-self, kirk-etal-2023-past, zhu2024principledreinforcementlearninghuman}. %The framework illustration is shown in Figure \ref{RL}.

\textbf{Supervised Fine-tuning}

\begin{figure}[htbp]
\includegraphics[width=0.48\textwidth]{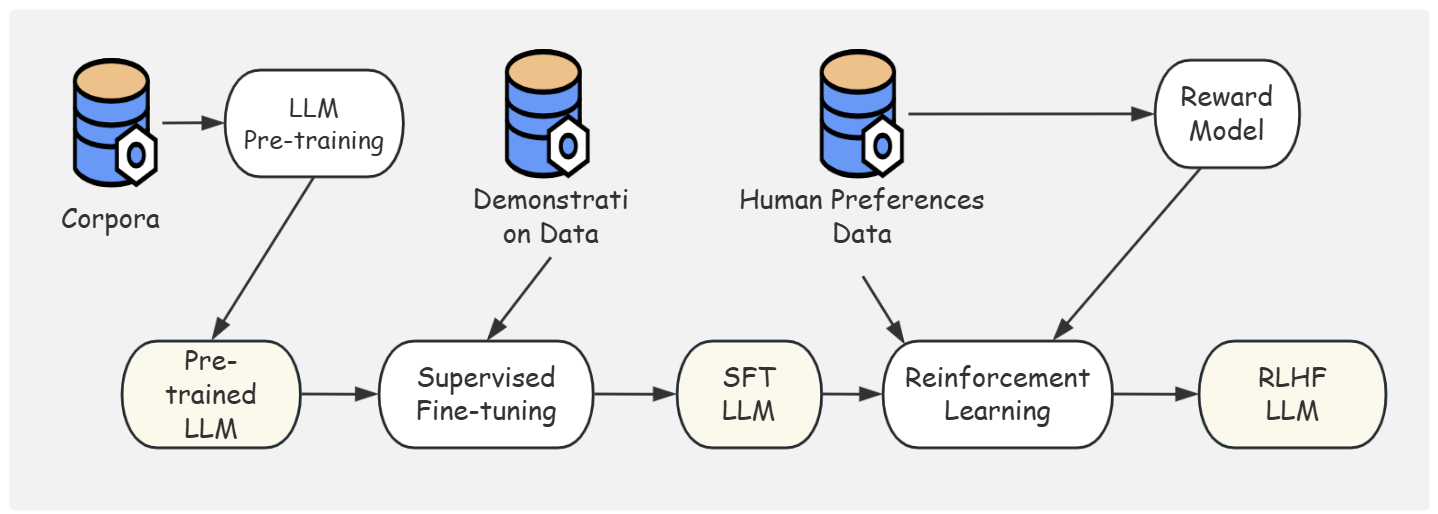} 
    \caption{An Illustration for Supervised Fine-tuning} % 
    \label{SFT} 
\end{figure}

Another way to promote human-agent agreement is Supervised Fine-tuning (SFT) illustrated  in Figure \ref{SFT} \cite{dong2023raftrewardrankedfinetuning, alpaca2023}, which compares the loss between LLMs' outputs and labelled datasets to update the model.
%as the framework illustration shown in Figure \ref{SFT}. 
These manual-annotated preference data mainly encompass human-written instruction-response pairs \cite{taori2023alpaca, ding-etal-2023-enhancing} and query-form preferences \cite{guo2024controllablepreferenceoptimizationcontrollable}. 
%Instruction-driven fine-tuning SFT is also known as Instruction-finetuning (IFT), which is mostly used in static tasks. 
For example, Instruction-finetuning (IFT), a form of instruction-driven SFT, is primarily used for static tasks. In contrast, preference labelling is usually adopted to capture users' personalised subtle preferences, and is mostly used in dynamic interactions. 
Examples of SFT include Stanford Alpaca \cite{alpaca2023} and AlpaGasus \cite{chen2024alpagasustrainingbetteralpaca},  demonstrating how their IFT fine-tuning leads to better instruction-following abilities. InstructGPT \cite{ouyang2022traininglanguagemodelsfollow} combines IFT with preference learning.
% it starts with SFT to fine-tune LLMs using a human-written prompt dataset, then uses labeler to rank multiple outputs of the fine-tuned model based on their preferences to train a reward model, which subsequently guides the fine-tuning of the 
Furthermore, frameworks like LIMA \cite{zhou2023limaalignment} and PRELUDE \cite{gao2024aligningllmagentslearning} introduce new angles to agreement fine-tuning by aligning user preferences through high-quality prompt-response pairs, learning users' latent preferences from dialogues and edit losses, rather than directly fine-tuning the pre-trained model. Also, 
\cite{yuan2024advancingllmreasoninggeneralists} introduces the Preference Tree, based on the ULTRAINTERACT dataset, 
% consisting of right and wrong reasoning chains, multi-turn interactions and preference data. 
enabling offline fine-tuning of LLMs via SFT by learning preferred reasoning paths. 
% in the Preference Tree 
% offline.

\begin{figure}[htbp]
\includegraphics[width=0.47\textwidth]{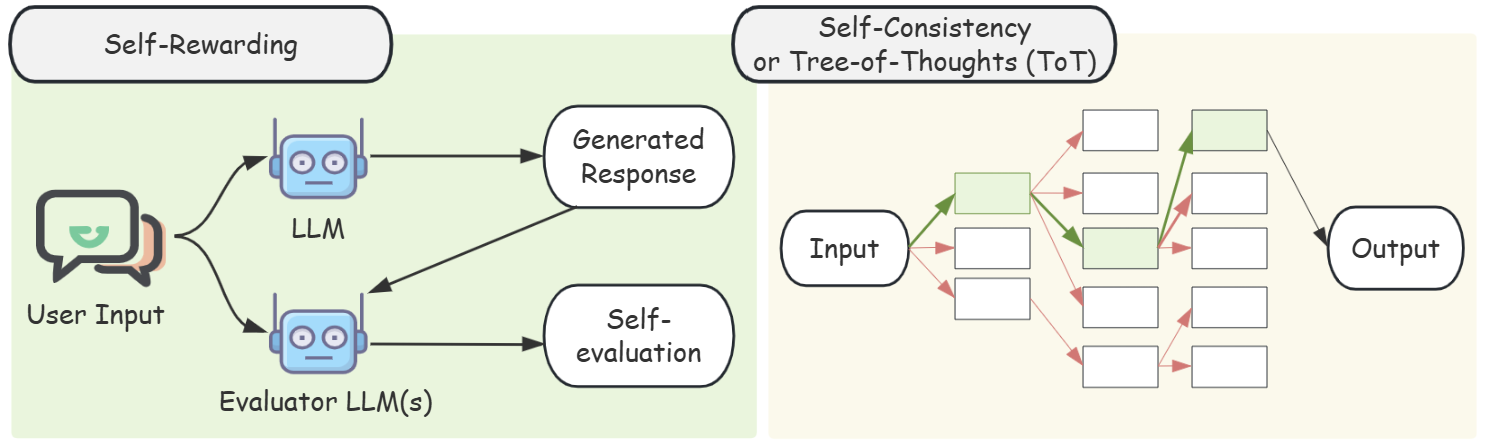} 
    \caption{Framework of Self-improvement} % 
    \label{Self-Improvement} 
\end{figure}

% \textbf{direct Human Feedback}
% \subsection{Single Agent Alignment}
\textbf{Self-improvement}
Inductive biases are used to refine agreement iteratively by self-improvement, as the framework illustration shown in Figure \ref{Self-Improvement}. Self-consistency \cite{wang2023selfconsistencyimproveschainthought} uses Chain-of-Thought (COT) \cite{wei2022chain} and Tree-of-thought (TOT) \cite{yao2024tree} to generate multiple reasoning paths and marginalise the response with the highest consistency when decoding to improve output quality. Based on this, Self-improve \cite{huang2022largelanguagemodelsselfimprove} chooses high-confidence inference paths as training samples to fine-tune more consistent models. 
%to become . 
SAIL \cite{ding2024sailselfimprovingefficientonline} utilize bi-level optimization, combining SFT and online RLHF 
%underpinned by 
to reduce the reliance on human annotated preferences. Self-rewarding \cite{yuan2024selfrewardinglanguagemodels} shows LLMs can improve preferences by judging their own answers. Based on this, Meta-Judge \cite{wu2024metarewardinglanguagemodelsselfimproving} add a meta-judging stage to optimist its judgement skills unsupervisedly. 

\iffalse
\textbf{Our Perspectives:}
Despite the availability of the aforementioned technologies to facilitate human-agent agreement, the robustness of these technologies still merits further investigation.
Some alignment-breaking attacks have been implemented on LLMs by role-play \cite{shen2024donowcharacterizingevaluating} or other designed response requirements to lead to jailbreak. Robustly Aligned LLM (RA-LLM) \cite{cao2024defendingalignmentbreakingattacksrobustly} build an alignment check function based on the assumption that LLMs have been aligned and are possible to reject malicious instructions, ensuring models' alignment when facing adversarial samples.
Behaviour Expectation Bounds \cite{wolf2024fundamentallimitationsalignmentlarge} is based on the assumption that LLMs can be divided into well/ill-behaved parts. LLMs' robustness under adversarial prompts can be ensured by setting bounds for behaviour weight, distinguishability, and similarity.

\fi

\subsection{Agent to Agent Agreement}
In a multi-agent system, agreement manifests as an agent's capability to accurately process other agents' intent, information, and output for informed collective decision-making \cite{10720863}. This section examines existing agreement mechanisms across heterogeneous agents.
% When we consider a multi-agent system, 
% % For multiagent systems, where agents usually keep interdisciplinary knowledge, U
% agreement usually refers to an agent's ability to accurately interpret and process other agents' intent, information, or output, to make more informed decisions together \cite{10720863}.
% % In reality, LLM-MAS are more universal due to their division of labour and high efficiency. 
% % Human-agent alignment focuses on how LLM agent understands human values, where human's role can be replaced by other agents alternatively \cite{akyürek2023rl4fgeneratingnaturallanguage}. Therefore, I
% In this section, we will look into how agents reach agreement.%, or how to achieve agreement among different sectors. % with humans through collaboration.

\begin{figure}[htbp]
\includegraphics[width=0.47\textwidth]{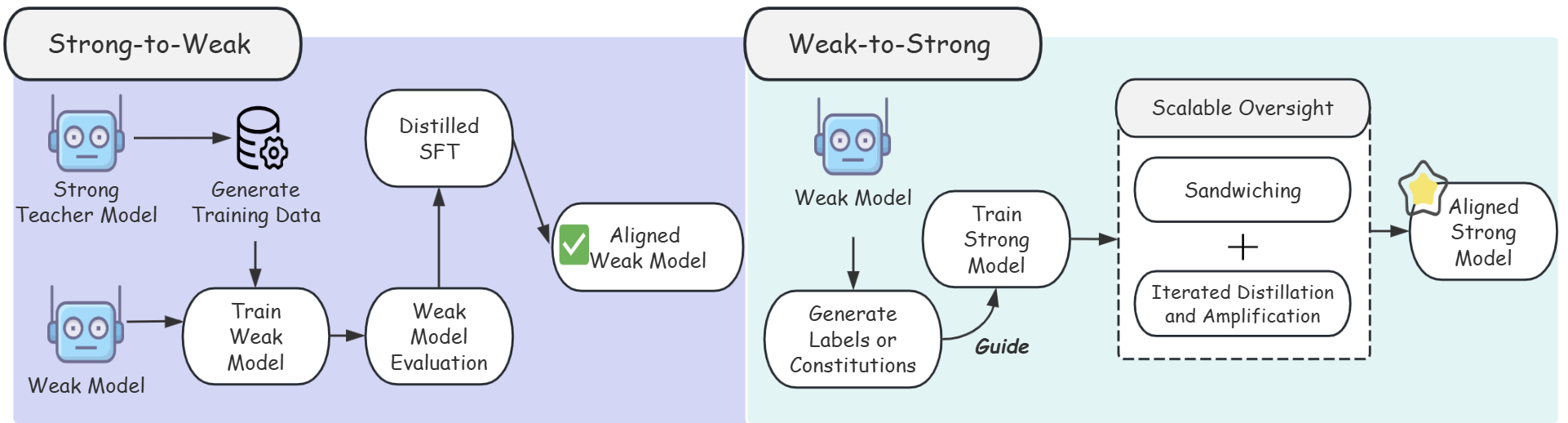} 
    \caption{Cross-Model Agreement Frameworks} % 
    \label{Cross-Model Agreement} 
\end{figure}

\textbf{Cross-Model Agreement}
There are two directions as shown in Figure \ref{Cross-Model Agreement}: One is Strong-to-weak. An aligned stronger teacher model generates training data for a weak model to learn behaviours, including response pairs \cite{xu2024wizardlm, taori2023alpaca, peng2023instructiontuninggpt4} and preferences \cite{cui2024ultrafeedbackboostinglanguagemodels}. For example, Zephyr \cite{tunstall2023zephyrdirectdistillationlm} 
% , following the structure of InstructGPT \cite{ouyang2022traininglanguagemodelsfollow} we mentioned in the SFT section, 
fine-tunes smaller LLMs through distilled SFT (dSFT). Before the last step DPO, the teacher LLM judge the smaller models' output as labellers instead of humans.
Another is Weak-to-strong. SAMI \cite{fränken2024selfsupervisedalignmentmutualinformation} writes constitutions using weak institution-fintuned models to avoid over-reliance on strong models. In \cite{burns2023weaktostronggeneralizationelicitingstrong}, weak teacher models are trained on ground truth by fine-tuning pre-trained models, which generate labels for strong student models. 
% This paradigm can compare scalable oversight methods \cite{10.5555/3495724.3495977}, such as Sandwiching \cite{bowman2022measuringprogressscalableoversight} and Iterated Distillation and Amplification (IDA) \cite{christiano2018supervisingstronglearnersamplifying}, which consider how humans supervise LLMs beyond human abilities. 
Considering the correlation of agents’ behaviours in collaboration, mutual information (MI) is also used to optimise cross-model agreement. A multi-agent reinforcement learning (MARL) method, Progressive Mutual Information Collaboration (PMIC) \cite{li2023pmicimprovingmultiagentreinforcement}, set the criterion that the MI of superior behaviours should be maximised and the MI of inferior ones should be minimised.

\begin{figure}[htbp]
\includegraphics[width=0.47\textwidth]{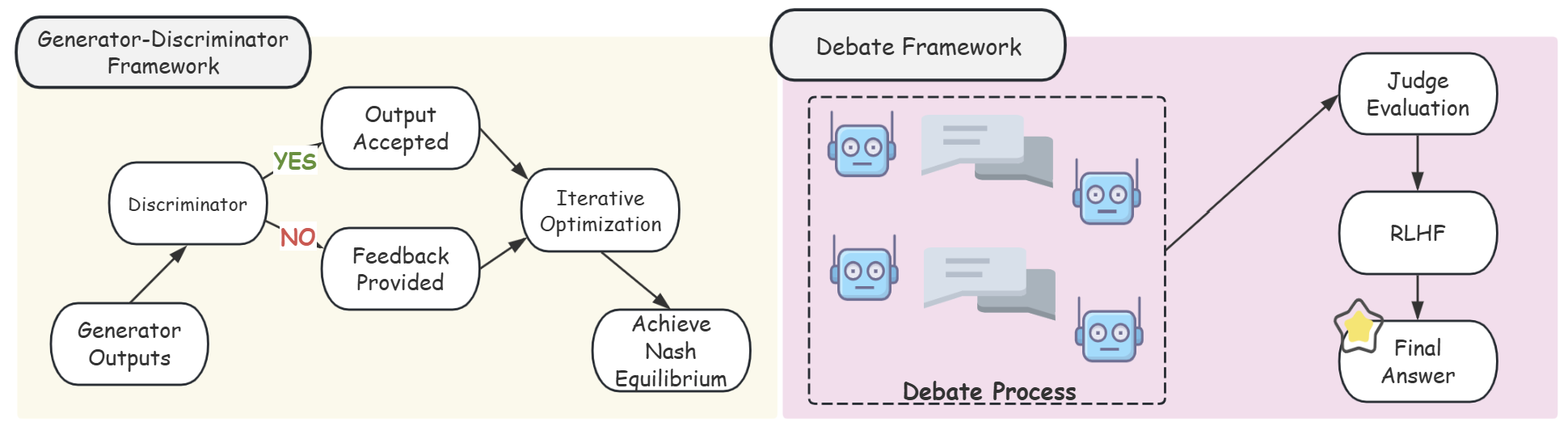} 
    \caption{Adversarial Self-Play and Debate Frameworks} % 
    \label{Debate} 
\end{figure}

\textbf{Debate and Adversarial Self-Play}
Debate normally exploits adversarial dynamics to refine agreement in a MAS, especially for an interdisciplinary MAS. There are two types: Generator-Discriminator and Debate, as shown in Figure \ref{Debate}. In the Generator-Discriminator framework, the generator generates the response, and the discriminator judges the quality. CONSENSUS GAME \cite{jacob2023consensusgamelanguagemodel} enhances agreement between a Generator and a Discriminator by iteratively refining their policies to minimize regret and reach a regularized Nash equilibrium.
With the Debate Framework, a debate process is simulated to improve the models' reasoning and agreement from strong opponents' perspectives. During the \cite{irving2018aisafetydebate}, Supervised pre-trained models play as debaters to generate arguments withstanding scrutiny, and RLHF is used to achieve a Nash equilibrium, enhancing agents' agreement with human expectations.

% PROVER-VERIFIER GAMES \cite{kirchner2024proververifiergamesimprovelegibility} achieve a Stackelberg equilibrium through Checkability Training, which can also help instruct humans to align with superhuman models. Stable agreement \cite{liu2023trainingsociallyalignedlanguage} combines adversarial training and consistency constraints with action space limitation to mitigate potential agreement instabilities, especially for out-of-distribution (OOD) inputs and adversarial prompts.

\begin{figure}[htbp]
\includegraphics[width=0.46\textwidth]{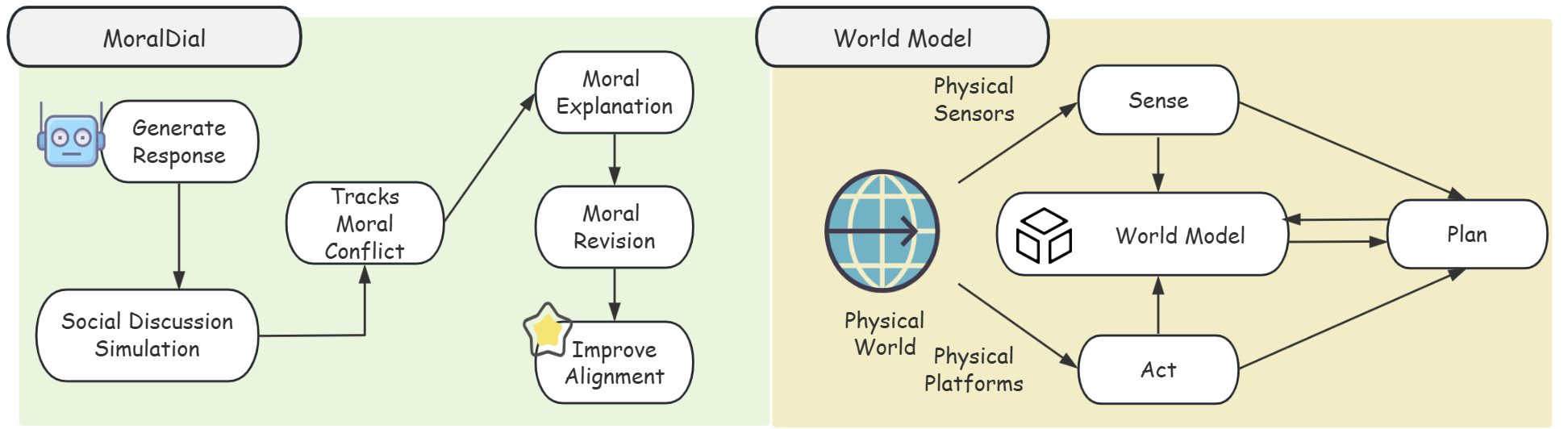} 
    \caption{Environment Feedback Frameworks} % 
    \label{Environment Feedback} 
\end{figure}

\textbf{Environment Feedback}
To achieve interdisciplinary agreement, a large amount of multimodal background knowledge is needed to build a World Model \cite{lecun2022path} for independent tasks and different roles, constituting a basis for \textit{common sense}. The agents' states and actions are the input, and the World Model provides multiple possible state predictions, such as state transition probabilities and relative rewards \cite{hu2023languagemodelsagentmodels}. The agents will find the strategy with the lowest estimated cost in the World over the long run under the \textit{common sense}.
Environment-driven tasks can also incorporate external tools and social simulations instead of purely manual annotation to expand agreement beyond language-based interactions to multimodal and task-specific applications.
Study like MoralDial \cite{sun2023moraldialframeworktrainevaluate} simulates social discussions between agents and the environment, improving the model's performance in moral answering, explanation, and revision, as shown in Figure \ref{Environment Feedback}.

\subsection{Agreement Evaluation}
% LLMs introduce unpredictability to agents' understanding because of their stochastic nature and synchronicity of knowledge. Therefore, we need mechanisms to measure whether the extent of the Agreement is acceptable for a multi-agent system. 
To effectively achieve and evaluate global agreement in a multi-agent system, dedicated evaluation methods to measure whether the extent of the agreement is acceptable for a MAS are essential.
The MAgIC \cite{xu2024magicinvestigationlargelanguage} brings forward metrics to evaluate capabilities within a MAS, where the Cooperation and the Coordination calculate the proportion of successful cases that achieve common goals compared with benchmarks. 
\cite{li2023quantifyingimpactlargelanguage} uses differences in opinions between individuals or groups to describe consistency, and uses the time for the difference in opinions between individuals decreasing to a threshold and standard deviation of group opinions to represent convergence. 
\cite{fung2024trustbasedconsensusmultiagentreinforcement, decerqueira2024trustaiagentsexperimental} introduce Trust Scores to evaluate how much an agent trusts others, which affects 
%the cooperation and coordination, and thus 
consensus in discussions. Each agent maintains a binary trust score for its Neighbours and updates the score based on others' behaviours in interactions.
Consensus is also measured by the degree of agreement of agents' final states after multiple rounds of negotiation. \cite{chen2025multiagentconsensusseekinglarge} believe the ultimate output represents a systematic consensus, so the consensus can be quantified by measuring the deviation by variance. 
Semantic Similarity \cite{xu2024reasoningcomparisonllmenhancedsemantic, aynetdinov2024semscoreautomatedevaluationinstructiontuned} is also used to assess the level of agreement among agents during their optimization process.
%-- the foundation of achieving a global agreement.

% Overall, this section reviews state-of-the-art methods for enhancing agent-to-human and agent-to-agent alignment (as shown in Figures \ref{humanagentalign} and \ref{agentagentalign}), achieving greater scalability and reducing the need for annotations. Additionally, several metrics for measuring understanding consistency and consensus are discussed, ensuring acceptable agreement.
% While some experiments have demonstrated positive alignment effectiveness on certain benchmarks, vulnerabilities remain in this area, particularly concerning performance optimization and robustness to out-of-distribution (OOD) data in weak-to-strong alignment scenarios. This paper will further examine these challenges and potential research directions for multiagent systems to achieve Agreement in detail in following sections.

% \section{Evaluation in LLM-empowered Multiagent System}
% \subsection{Single Agent Performance}

% \subsection{Agent-Agent Alignment}

% \subsection{}

\section{Uncertainty in LLM-MAS} 
With the shift from single-agent planning to multi-agent collaboration, uncertainty management becomes a crucial external challenge for ensuring a responsible LLM-MAS. This requires effective traceability, probabilistic guarantees, and strategic utilization of uncertainty across all system components. This section explores how uncertainty quantification techniques enhance AI agents and evaluation metrics, facilitating the transition to multi-agent setups and fostering more robust, reliable MAS for responsible decision-making.

\subsection{Uncertainty in AI Agents System}
Despite the widespread deployment of LLM across various domains, 
% including medical diagnostic assistance~\cite{wei2024medaide,abbasian2023conversational} and physical applications such as robotic manipulation and navigation~\cite{ahn2022can, mandi2024roco}, 
the explicit consideration of uncertainty in LLM-empowered agents remains relatively unexplored. 
When we analyse an AI-agent system by breaking it down into individual components, it transforms into a multi-component system. Therefore, 
%in this section, 
we firstly focus on the core components that influence the AI agent's uncertainty, e.g. memory management, and strategic planning.
% Ensuring the safety and trustworthiness of these systems is critical, as erroneous outputs can potentially result in severe or even catastrophic consequences. 
% To develop responsible and reliable LLM-based agent and multi-agent systems, it is essential to systematically investigate uncertainty in some key components such as memory management, strategic planning, and self-evaluation, with a particular focus on integrating uncertainty quantification methodologies. 
% Addressing these aspects will not only enhance the robustness of LLM-driven agents but also foster greater confidence in their deployment across high-stakes applications.

\textbf{Memory} 
Retrieval-Augmented Generation (RAG) enhances LLMs by integrating external, up-to-date, domain-specific knowledge, improving factual accuracy and reducing hallucinations without extensive retraining. However, not all retrieved sources equally influence decision-making. To address this, an attention-based uncertainty quantification
% approach has been proposed
~\cite{duan2024shifting}
% which estimates uncertainty by 
analyzes variance in attention weights across retrieved sources to estimate uncertainty. Similarly, LUQ~\cite{zhang2024luq} uses an ensemble-based approach to re-rank documents and adjust verbosity based on confidence.
% , ensuring improved factual accuracy in long-form generation. Additionally, consistency between retrieved information and generated content is essential for reliability. 
Xu et al.~\cite{xu2024unsupervised} introduce a self-consistency mechanism,
% to quantify uncertainty by 
comparing retrieved evidence with generated outputs 
% , using discrepancies as uncertainty indicators 
to refine both retrieval and generation, ultimately improving the model's knowledge representation and reducing hallucinations.
% Retrieval-Augmented Generation (RAG) plays a critical role in enhancing LLMs by dynamically integrating external, up-to-date, and domain-specific knowledge, thereby improving factual accuracy and mitigating hallucinations without %necessitating 
% extensive retraining. However, not all retrieved %information 
% sources contribute equally to the decision-making process. To address this issue, an attention-based uncertainty quantification approach has been proposed~\cite{duan2024shifting}, which estimates uncertainty by analysing the variance in attention weights across multiple retrieved sources. 
% % This approach prioritizes high-confidence responses while filtering out less reliable outputs, thereby enhancing the robustness of free-form text generation. 
% Similarly, the LUQ~\cite{zhang2024luq}
% %framework
% employs an ensemble-based approach to re-rank retrieved documents and adjust response verbosity based on confidence levels, ensuring improved factual accuracy in long-form text generation. Moreover, consistency between retrieved information and model-generated content is crucial for maintaining reliability. Xu et al.~\cite{xu2024unsupervised} introduce a self-consistency mechanism to quantify uncertainty by comparing retrieved evidence with generated outputs. Discrepancies identified among sources serve as uncertainty indicators, which are leveraged to refine both retrieval and generation processes, ultimately improving the model's internal knowledge representation and reducing hallucinations in RAG-based systems.

\textbf{Planning} Planning is another essential component for LLM-based agents as it enables structured decision-making by decomposing complex tasks into manageable steps.
% , ensuring coherence and goal-oriented execution over extended interactions. 
However, planning remains the most uncertain aspect in a stochastic environment. To 
address uncertainty in stochastic environments; studies focus on improving efficiency and reliability. Tsai et al.~\cite{tsai2024efficient} fine-tunes Mistral-7B to predict prompt-action compatibility, 
% scores between user prompts and generated actions, 
using conformal prediction to identify the most probable actions. 
To assess the need for human evaluation, Ren et al. \cite{ren2023robots} introduce KnowNo, a method that evaluates token probabilities for next actions. 
Building on this, IntroPlan \cite{liang2024introspective} incorporates introspective planning, refining 
% the model's ability to generate 
prediction sets with tighter confidence bounds, reducing 
% reliance on 
human intervention and enhancing 
% decision-making efficiency and 
autonomy.
% Planning is another essential component for LLM-based agents as it enables structured decision-making by decomposing complex tasks into manageable steps, ensuring coherence and goal-oriented execution over extended interactions. However, planning is the most uncertain part of an agent in a stochastic environment. To enhance decision-making efficiency and reliability, the study~\cite{tsai2024efficient} fine-tunes the Mistral-7B LLM to predict compatibility scores between a user-provided prompt and generated actions. By applying conformal prediction, the most probable actions are identified and presented to the user for selection. To assess the necessity of human evaluation in decision-making, Ren et al.~\cite{ren2023robots} introduce KnowNo, a method that evaluates the token probability assigned to each potential next action. Building upon this approach, IntroPlan~\cite{liang2024introspective} extends KnowNo by incorporating introspective planning, which enhances the model's ability to generate prediction sets with tighter confidence bounds. This refinement reduces the reliance on human intervention, thereby improving the efficiency and autonomy of the decision-making process.

% \paragraph{Rationality}

\subsection{Uncertainty in Agents Interaction}
While uncertainty quantification in LLM-MAS has been explored, existing methods typically assess uncertainty at individual instances, overlooking prior interaction history. Real-world applications, like autonomous medical assistants \cite{li2024mediq, savage2024large}, often involve multi-instance interactions, where responses depend on accumulated information from previous exchanges \cite{chen2024llmarena, pan2024agentcoord}. 
% For instance, in an autonomous medical assistant, tasks such as patient history collection and medical image analysis rely on prior actions, requiring a collective approach to uncertainty measurement \cite{li2024mediq, savage2024large}. 
% In these scenarios, uncertainty should be evaluated across interactions, not in isolation. 
In multi-agent settings, 
% an effective uncertainty quantification method should ensure consistent recall of correct answers across varied question formulations. For example, 
methods like DiverseAgentEntropy~\cite{feng2024diverseagententropy} assess uncertainty by evaluating 
% evaluates an LLM's uncertainty by assessing 
factual parametric knowledge in a black-box setting, providing a more accurate prediction and helping detect hallucinations. It further reveals that existing models often fail to consistently retrieve correct answers across diverse question formulations, even when the correct answer is known. Moreover, 
% uncertainty can result in a negative impact on multi-agent interactions when agents 
failure to express uncertainty explicitly can 
% levels, leading to the potential 
misguidance other agents \cite{liang2023encouraging, burton2024large}. DebUnc~\cite{yoffe2024debunc} tackles this issue by incorporating confidence metrics throughout the entire interaction, improving the clarity and reliability of agent communication. 
It adapts the LLM attention mechanism to adjust token weights based on confidence levels and uses textual prompts to convey confidence more effectively. 
\vspace{-10pt}
\subsection{Uncertainty Evaluation}

From an agent-monitoring perspective, the performance of LLM-MAS can be assessed using statistical metrics, through human-in-the-loop verification, or a combination of both. Ideally, to minimize human intervention and enhance the efficiency of responsible agent systems, only outputs identified as uncertain should be deferred to an auxiliary system or human experts for further evaluation.

\textbf{Statistical Analysis} Uncertainty estimation in LLMs can be broadly categorized into single-inference and multi-inference approaches. Single-inference estimation 
% quantifies uncertainty based on a single output, utilizing 
use token log probabilities
% . For instance, in free-form language generation tasks, 
with logit values 
% can 
partially 
capture inherent uncertainty
% , with studies showing that models can adjust their responses based on this uncertainty to improve performance
~\cite{yang2023improving}, while conformal prediction~\cite{ren2023robots} further quantifies confidence
% , ensuring the agent operates within a 
for predefined success probabilities~\cite{ren2023robots}. In contrast, multi-inference estimation evaluates uncertainty across multiple outputs, bypassing token-level details.
% the need for individual token log probabilities. 
Intuitively, if a model has effectively learned a concept, its generated samples should exhibit semantic equivalence. Methods like Semantic entropy~\cite{farquhar2024detecting} 
% leverage entropy-based estimators to 
detects confabulations (arbitrary and incorrect generations) by measuring uncertainty at the semantic level, 
% enabling task-agnostic generalization. Additionally,
and spectral clustering~\cite{lin2023generating} quantifies uncertainty by analyzing semantic dispersion in multiple responses, providing a robust estimate without accessing 
% the model's 
internal parameters.
\textbf{Human-in-the-loop} 
Setting an uncertainty threshold helps identify potential errors and delegate high-risk cases to external systems or human experts, with outcomes exceeding the threshold flagged for reassessment. For example, KnowLoop framework~\cite{zheng2024evaluating} uses entropy-based measures 
% to quantify unpredictability, enabling 
for failure detection and human intervention in LLM-based task planning.
% when uncertainty surpasses predefined thresholds
Similarly, UALA~\cite{han2024towards} integrates uncertainty quantification into its workflow, using metrics like maximum or mean uncertainty 
% within a calibration set 
to identify knowledge gaps,  
% or errors. Exceeding this threshold 
prompting the agent to seek clarification. 
% additional information or clarification for the identified knowledge gap. 
These mechanisms enhance the robustness and adaptability of LLM-based systems, reducing risks from erroneous outputs and improving reliability across diverse applications.
Despite recent progress in uncertainty quantification, LLM-MAS still lacks rigorous uncertainty measures that both incorporate traceable agent interaction histories and establish verifiable statistical bounds, which is a critical requirement for developing responsible LLM-MAS frameworks.

% \section{Systematic Design of Responsible LLM-MAS Framework}
\section{Responsible LLM-MAS Framework}
% To build a safer LLM-MAS demands interdisciplinary perspectives as safety mechanisms vary significantly across domains
Building a responsible LLM-MAS inherently requires interdisciplinary perspectives, as safety mechanisms vary across domains \cite{gao2024large}. For instance, majority voting works for a content recommendation but fails in healthcare, where minority expert opinions are critical. These domain-specific considerations can be integrated into LLM-MAS through structured prompting mechanisms, incorporating predefined rules, knowledge graphs, or domain ontologies. Meanwhile, trustworthy specifications are enforced via validation rules and operational constraints \cite{handler2023balancing}. This structured integration guides LLMs' behavior according to domain expertise and regulatory requirements, ensuring safety while preserving the systems' responsibility.

Another crucial aspect of responsible LLM-MAS design lies in establishing quantifiable guarantee metrics, involving 
% that provide essential measurements of system reliability through comprehensive 
agreement evaluation and uncertainty quantification. The agreement dimension involves multiple levels of assessment, including but not limited to: consensus among agent decisions, policy alignment, goal consistency, etc. Additionally, system-wide considerations such as communication protocol compliance, privacy information propagation, and temporal synchronization constraints must be carefully evaluated across the multiagent network \cite{he2025llm}. Meanwhile, uncertainty quantification operates at both system and agent levels, addressing various aspects such as knowledge confidence assessment, decision reliability estimation, and environmental state prediction, among others. 
These metrics, with probabilistic bounds, ensure operational risks stay within acceptable margins 
%, preventing catastrophic failures 
\cite{nikolaidis2004comparison,hsu2023safety}. These quantifiable guarantee metrics not only enable objective evaluation of system trustworthiness and performance but also serve as the foundation for building robust monitoring mechanisms.
\begin{figure}[htbp]
    \setlength{\abovecaptionskip}{-1pt} 
    \setlength{\belowcaptionskip}{-1pt}
    \centering
\includegraphics[width=1.0\columnwidth]{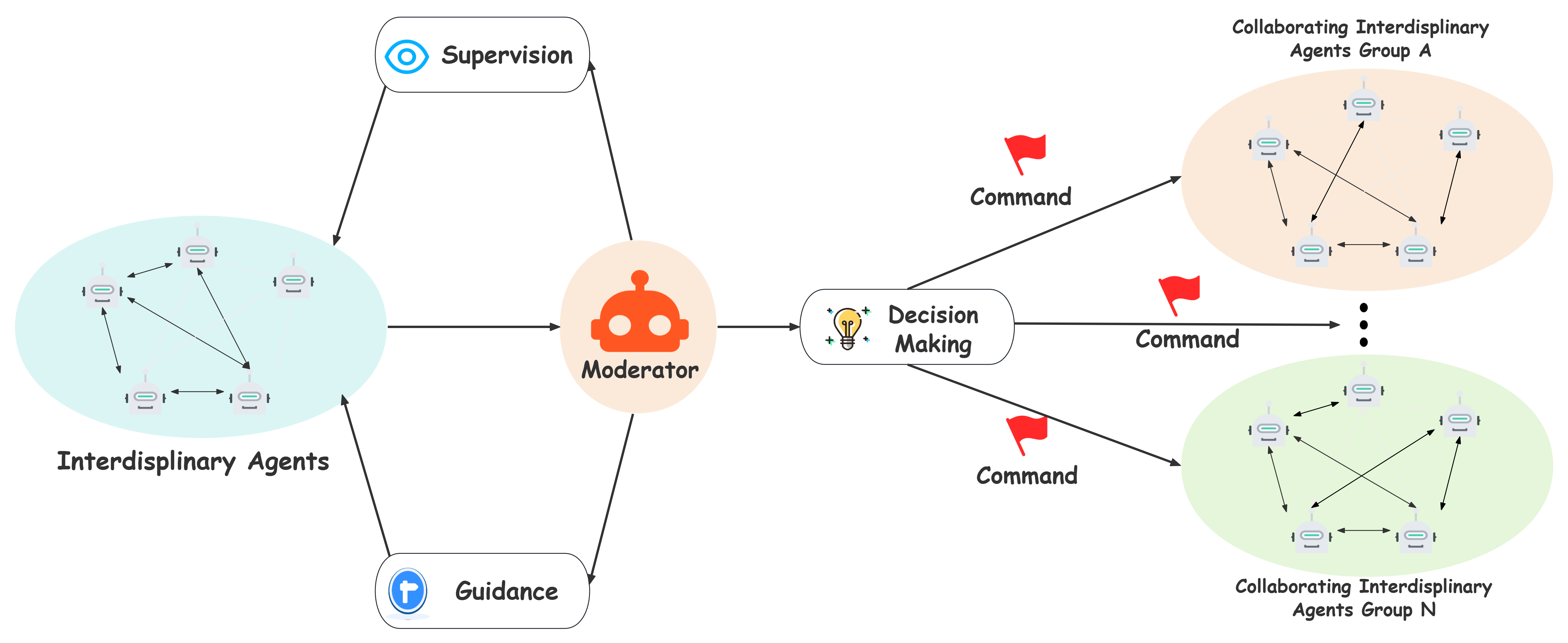} 
    \caption{Illustration of Responsible LLM-MAS Framework} % 
    \label{safeframework} 
\end{figure}

% Leveraging these quantifiable guarantees, we can design 
A moderator, 
% which serves as a core agent by 
integrating symbolic rules with formal verification, can manage the system rigorously, as shown in Figure \ref{safeframework}. Unlike LLM-as-judge approaches that lack formal guarantees, this moderator should employ these metrics to validate results and possess dynamic recovery strategies to solve discrepancies \cite{benner2021you}. 
% When detecting agreement discrepancy, the moderator can employ a multi-faceted recovery strategy including state restoration, rollback procedures, and knowledge manipulation methods. 
These mechanisms ensure system resilience by facilitating the re-establishment of inter-agent agreement through controlled recovery processes, while maintaining operational efficiency. The moderator's ability to provide interpretable guarantees stems from its verifiable metric-based assessment and human-involved design while adapting to dynamic situations through its hybrid architecture, combining the rigour of formal methods with the flexibility of LLM-based reasoning.

% interdisciplinary consideration

% quantifiable guarantee metric (agreement \& uncertainty)

% super monitor agent
% \jinwei{Should we add this section?}

% \subsection{Neural-Symbolic Approach?}

% \textbf{Our Position:}
% \section{Defence/Optimisation in LLM-empowered Multiagent System}

\section{Conclusion} 
% This paper introduces a principled framework for responsible LLM-MAS, addressing three fundamental aspects: comprehensive agreement mechanisms for inter-agent alignment, quantifiable uncertainty guarantees across agent interactions, and system-level oversight and intervention. Through analysing challenges in modern LLM-MAS, we propose potential technical perspectives spanning AI provenance, neural-symbolic methods, formal verification, and other promising domains for system dependability. This work lays the foundation for developing robust, trustworthy LLM-MAS deployments, while opening new research directions in uncertainty quantification, agreement evaluation, and dynamic adaptation mechanisms for complex multi-agent interactions.

This 
position paper advocates for a responsible framework for building LLM-MAS beyond the current solutions, which only offer the simplest mechanisms based on predefined rules. 
LLM-MAS are highly complex due to their role of managing interaction among agents, uncertainty from environments, and human-involved factors. 
A responsible framework, supported by multidisciplinary agents and expert moderators, can fully consider and manage the complexity and provide assurance to the final product.

\newpage
% jinwei modified for arXiv
\clearpage
\bibliography{example_paper}
\bibliographystyle{icml2025}

%%%%%%%%%%%%%%%%%%%%%%%%%%%%%%%%%%%%%%%%%%%%%%%%%%%%%%%%%%%%%%%%%%%%%%%%%%%%%%%
%%%%%%%%%%%%%%%%%%%%%%%%%%%%%%%%%%%%%%%%%%%%%%%%%%%%%%%%%%%%%%%%%%%%%%%%%%%%%%%
% APPENDIX
%%%%%%%%%%%%%%%%%%%%%%%%%%%%%%%%%%%%%%%%%%%%%%%%%%%%%%%%%%%%%%%%%%%%%%%%%%%%%%%
%%%%%%%%%%%%%%%%%%%%%%%%%%%%%%%%%%%%%%%%%%%%%%%%%%%%%%%%%%%%%%%%%%%%%%%%%%%%%%%
%%%%%%%%%%%%%%%%%%%%%%%%%%%%%%%%%%%%%%%%%%%%%%%%%%%%%%%%%%%%%%%%%%%%%%%%%%%%%%%
%%%%%%%%%%%%%%%%%%%%%%%%%%%%%%%%%%%%%%%%%%%%%%%%%%%%%%%%%%%%%%%%%%%%%%%%%%%%%%%

\end{document}